\documentclass[utf8]{FrontiersinHarvard}
\usepackage{url,hyperref,lineno,microtype,subcaption}
\usepackage[onehalfspacing]{setspace}

\usepackage{changes} 

\def\keyFont{\fontsize{8}{11}\helveticabold }
\def\firstAuthorLast{Madanian {et~al.}} 
\def\Authors{H. Madanian\,$^{1,*}$, Y. Pfau-Kempf\,$^{2}$, R. Rice\,$^{3,4}$, T. Liu\,$^{5}$, T. Karlsson\,$^{6}$, S. Raptis\,$^{7}$, D. Turner\,$^{7}$ and J. Beedle\,$^{8}$}

\def\Address{$^{1}$CUA at NASA Goddard Space Flight Center, Greenbelt, MD, USA \\
$^{2}$University of Helsinki, Helsinki, Finland \\
$^{3}$University of Maryland at College Park, College Park, MD, USA \\
$^{4}$NASA Goddard Space Flight Center, Greenbelt, MD, USA \\
$^{5}$University of California Los Angeles, Los Angeles, CA, USA \\
$^{6}$KTH Royal Institute of Technology, Stockholm, Sweden \\
$^{7}$Johns Hopkins University Applied Physics Laboratory, Laurel, MD, USA \\
$^{8}$University of New Hampshire, Durham, NH, USA}

\def\corrAuthor{Hadi Madanian}
\def\corrEmail{hmadanian@gmail.com}

\begin{document}
\onecolumn
\firstpage{1}


\title[Sunward Flows in Magnetosheath]{Sunward Flows in the Magnetosheath Associated with Magnetic Pressure Gradient and Magnetosheath Expansion}

\author[\firstAuthorLast ]{\Authors}
\address{}
\correspondance{}
\extraAuth{}

\maketitle

\begin{abstract}
A density structure within the magnetic cloud of an interplanetary coronal mass ejection impacted Earth and caused significant perturbations in plasma boundaries. We describe the effects of this structure on the magnetosheath plasma downstream of the bow shock using spacecraft observations. During this event, the bow shock breathing motion is evident due to the changes in the upstream dynamic pressure. A magnetic enhancement forms in the inner magnetosheath and ahead of a plasma compression region. The structure has the characteristics of a fast magnetosonic shock wave, propagating earthward and perpendicular to the background magnetic field further accelerating the already heated magnetosheath plasma. Following these events, a sunward motion of the magnetosheath plasma is observed. Ion distributions show that both the high density core population as well as the high energy tail of the distribution have a sunward directed flow indicating that the sunward flows are caused by magnetic field line expansion in the very low $\beta$ magnetosheath plasma. Rarefaction effects and enhancement of the magnetic pressure in the magnetosheath result in magnetic pressure gradient forcing that drives the expansion of magnetosheath magnetic field lines. This picture is supported by a reasonable agreement between the estimated plasma accelerations and the magnetic pressure gradient force.
\tiny
 \keyFont{ \section{Keywords:} Shocks, Magnetosheath, Space Weather, Coronal Mass Ejections, Solar Wind, Space Plasmas} 
\end{abstract}




\section{Introduction} \label{sec:intro}

The magnetosheath region at Earth and other planetary systems stands between the upstream solar wind and the downstream magnetic obstacle (e.g., the magnetosphere or the magnetic pileup boundary). The magnetosheath region contains heated and compressed solar wind plasma that has been scattered and slowed down to subsonic speeds. Solar wind heating involves a variety of microphysical processes that are largely dependent on upstream plasma and shock parameters \citep{krasnoselskikh_dynamic_2013,burgess_microstructure_2016}. These include the bow shock inclination angle ($\theta_{Bn}$) with respect to the interplanetary magnetic field (IMF) and the upstream Mach number. The magnetosheath plasma downstream of quasi-parallel shocks, where $\theta_{Bn} < \sim 45^{\circ}$, is more turbulent than the quasi-perpendicular side of the bow shock. Such asymmetries can continue through the magnetosheath and be imposed on the magnetopause \citep{madanian_asymmetric_2022, gurchumelia_comparing_2022}. At supercritical shocks, heating and energy dissipation is partly through ion reflection \citep{schwartz_energy_2022}, the rate of which is dependent on the magnetic amplification at the shock and the magnetization \citep{madanian_drivers_2024}. Hot upstream ion populations with larger pitch angles are reflected easier upon encountering a magnetic boundary \citep{burgess_alpha_1989}. Heavy ions, such as alpha particles and singly charged helium ions, in a proton dominated solar wind plasma interact differently with the bow shock resulting in an unstable shock layer \citep{broll_mms_2018, madanian_interaction_2024}.

In addition to upstream effects, the magnetosheath plasma is driven by factors including the deflection pattern around Earth at the point of measurement and transient effects generated locally, or transported from downstream such as surface waves \citep{plaschke_magnetopause_2013,burkholder_complexity_2023}. The energy density of the magnetosheath plasma drives the magnetopause boundary stand off distance, an important parameter in space physics that determines the state of the magnetosphere and the magnetosphere-solar wind coupling rate. Empirical models relate the location of the magnetopause boundary to the dynamic pressure in the solar wind by using the hydrodynamic theory and assuming that flow pressure is entirely converted to thermal pressure in the magnetosheath \citep{chapman_new_1931}. Other models also include the IMF $B_z$ component as a proxy to consider the reconnection effects \citep{shue_magnetopause_1998}. The bow shock boundary distance in these models is simply scaled from the magnetopause based on the upstream Mach number in the solar wind \citep{farris_determining_1994}.

Magnetosheath plasma jets are known as periods of high dynamic pressure caused by either increases density or earthward flow velocity of the magnetosheath plasma \citep{kramer_jets_2025}. Magnetosheath jets are typically localized, constrained in size \citep{fatemi_unveiling_2024}, and are formed due to a variety of processes including foreshock effects, upstream discontinuities, and microphysical effects at the bow shock \citep{plaschke_jets_2018}. Discontinuities in the solar wind can also rattle the boundaries. The interaction of a tangential discontinuity (e.g., a density structure) with the bow shock and magnetosphere launches a fast mode magnetosonic shock wave through the magnetosheath \citep{maynard_cluster_2008, wu_magnetospheric_1993}. Another form of magnetic enhancement in the magnetosheath known as paramagnetic plasmoids can also form during the passage of upstream discontinuities \citep{karlsson_origin_2015}. Both the fast shocks and paramagnetic plasmoids are compressive structures. Upon encountering the magnetopause, the magnetosheath plasma is typically either deflected around the magnetosphere or enters the magnetosphere through reconnection. The ion plasma $\beta$ difference and the magnetic shear angle between the magnetosphere and the magnetosheath plasmas influence the reconnection rate at the magnetopause \hbox{\citep{phan_dependence_2010}}.

Sunward flows in the magnetosheath are rare. Some observational studies associate sunward flows with the magnetopause boundary motion in response to either a change in the upstream dynamic pressure, or due to indentation of the magnetopause boundary \citep{siscoe_sunward_1980, shue_anomalous_2009,archer_role_2014, zhou_magnetosheath_2024, farrugia_effects_2018}. As the magnetopause bounces outward, it drives the magnetosheath plasma with different $\beta$ and magnetic Reynold number with it, creating sunward flows. This process in in-situ spacecraft observations manifests as sunward plasma flows followed by a full or partial magnetopause crossing or the presence of the boundary layer plasma. The specific properties of the magnetosheath plasma have direct consequences on the reconnection rate and the amount of energy transfer at the magnetopause. As such, characterizing properties and dynamics of the magnetosheath is important in understanding the connected Sun-Earth system. In this paper, we investigate the properties and the underlying cause of sunward flows observed in the Earth's magnetosheath during a period of very low $\beta$ solar wind flow. The paper is organized as follows: analysis of in-situ observations of the solitary magnetic enhancement and sunward flows are described in Section \ref{sec:obs}, discussion and interpretation of results are provided in Section \ref{sec:disc}, and conclusions are provided in Section \ref{sec:conc}. Links to data sources are provided in Section \ref{sec:opres}.

\section{Data and Methods}
In this study, we use data from THEMIS (Time History of Events and Macroscale Interactions during Substorms) \citep{angelopoulos_themis_2008}, Cluster \citep{escoubet_cluster_2001}, (MMS) Magnetospheric Multiscale \citep{burch_magnetospheric_2016}, and Wind \citep{harten_design_1995} missions. The fortuitous configuration of these spacecraft on the dayside geospace allows for multi-point study of this event. The magnetic field data are from the magnetometer instruments (FGM) onboard the spacecraft. For THEMIS and MMS we use 16 Hz magnetic field data products, while for Cluster the data cadence is 5 Hz. The THEMIS ion data are from the reduced distributions of the electrostatic analyzer (ESA). All plasma moments by THEMIS spacecraft are recalculated from the returned distributions. The solar wind dynamic pressure is calculated from measurements by the Wind Three-Dimension Plasma (3DP) instrument \citep{lin_three-dimensional_1995}. 

\section{Observations and Analysis Results} \label{sec:obs}

\begin{figure}
\centering
\includegraphics[width=0.65\textwidth]{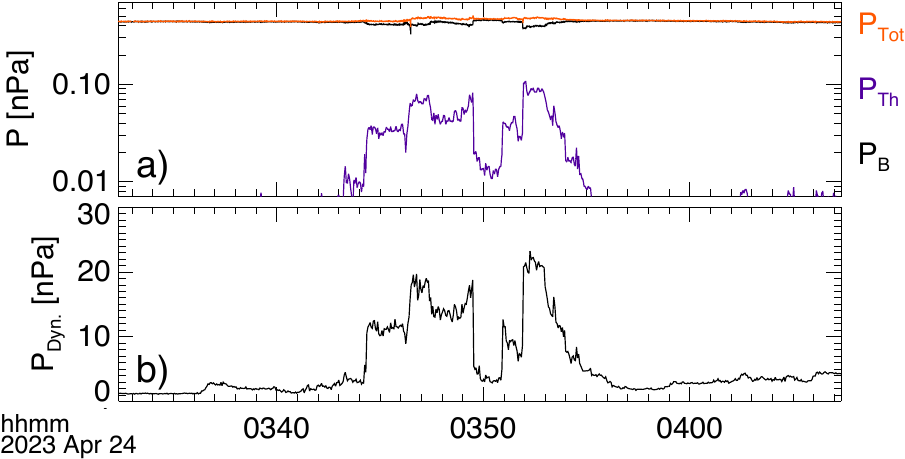}
\caption{Pressure terms associated with the density structure in the solar wind observed the solar wind monitor. The data are shifted in time to account for the travel time to the Earth's bow shock. Panel (a) shows the thermal pressure in blue, magnetic pressure in black, and the total pressure in red in logarithmic scale. Panel (b) shows the dynamic pressure.}
\label{fig:fig1_pressureterms}
\end{figure}

In this section we describe the impact of a density structure in the solar wind on the magnetosheath. The density structure is observed within the magnetic cloud of an ICME observed on 24 April 2023. The upstream solar wind plasma conditions and the interaction of the structure with the Earth's bow shock have been characterized in earlier studies \citep{madanian_interaction_2024}, and it has been shown that high abundances of protons, alpha particles, and singly charged helium ions exist within the density enhancement. This event caused significant geomagnetic activity and displacement of the bow shock and magnetopause from their nominal positions \citep{liu_global_2024}. Prior to the onset of the density peak, the solar wind plasma is dominated by the high magnetic pressure in the strong magnetic fields of the magnetic cloud (very low $\beta$ plasma). Figure \ref{fig:fig1_pressureterms} shows different pressure terms in the solar wind as measured by the solar wind monitor at Lagrange point 1 and shifted to the Earth's bow shock by a 42 minutes lag time. The second panel in this figure shows that during the density structure, high dynamic pressures ($P_{Dyn.}=\rho \cdot |V_{SW}|^2$, where $\rho$ is the mass density and $|V_{SW}|$ is the solar wind speed) are superimposed on top of the relatively high magnetic pressure solar wind flow.

\begin{figure}
\centering
\includegraphics[width=0.95\textwidth]{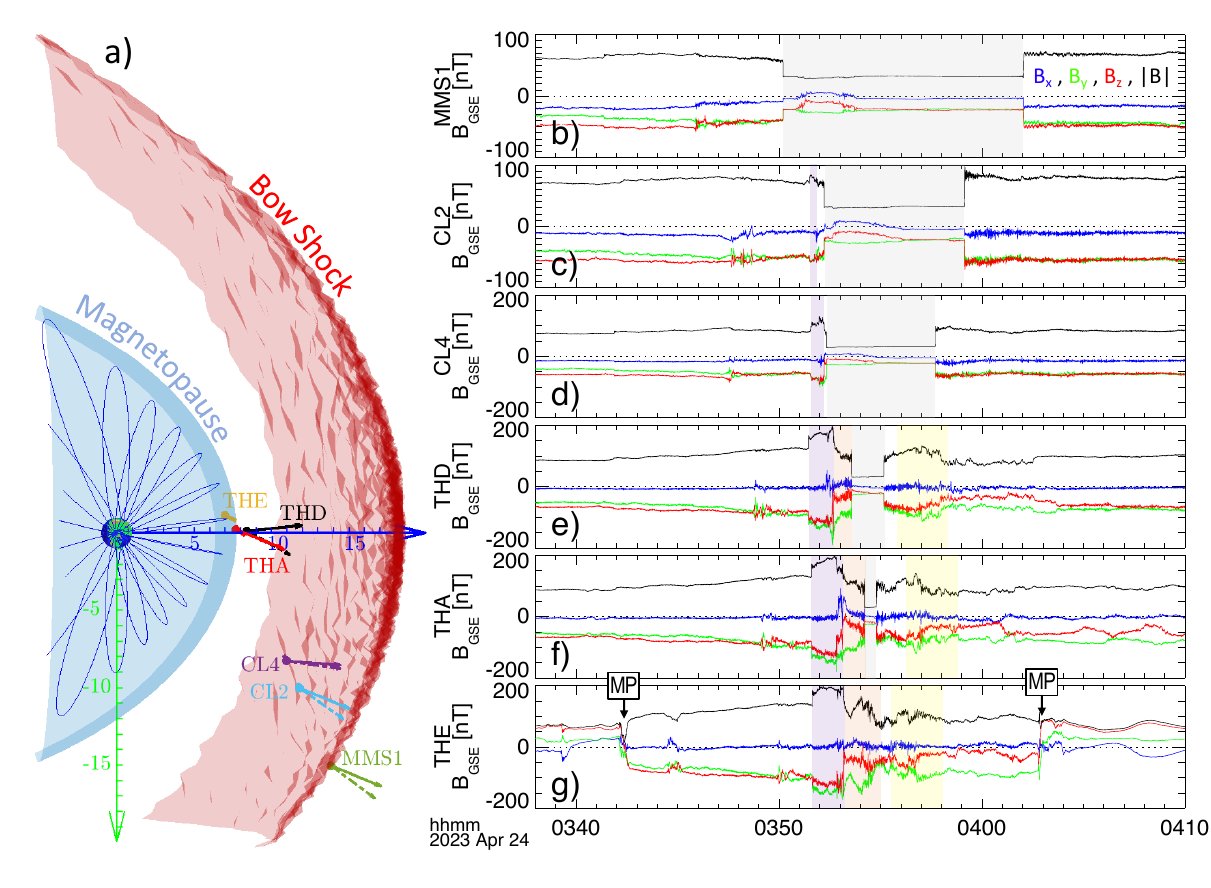}
\caption{Spacecraft positions and magnetic field measurements. The schematic on the left in panel (a) shows the projection of spacecraft positions on the $xy$ plane. The axes carry units of Earth radii (R\textsubscript{E}). Vector quantities in this figure and throughout the text are in the geocentric solar ecliptic (GSE) coordinates in which $+x$ is towards the sun, $+y$ is opposite to the planetary orbital motion, and $+z$ completes the right hand triple. The solid and dashed vectors originating from each spacecraft show the normal vector to the bow shock during the receding and expanding motion, respectively. Panels (b) through (g) show the magnetic field components and magnitude measured by each spacecraft. Shaded areas highlight the following, gray for the solar wind, purple for the fast magnetosonic shock, pink for the density pileup, and yellow for the sunward flow periods. Instances of magnetopause crossing by the TH-E spacecraft are marked by ``MP'' on the last panel.}
\label{fig:fig2_Boview}
\end{figure}

We analyze data from three constellations of spacecraft, namely THEMIS, Cluster, and MMS, positioned across the dayside magnetosheath during this event. Figure \ref{fig:fig2_Boview} shows the spacecraft positions with respect to the nominal bow shock and magnetopause boundaries. The conic section parameters are selected to match the MMS1 and THEMIS-E (TH-E) crossings of the bow shock and magnetopause, respectively. MMS spacecraft are initially inside the magnetosheath and close to the bow shock. We only show data from MMS1, since the four MMS spacecraft are in a close tetrahedron formation and make similar observations. TH-D, TH-A, and TH-E spacecraft are inside the magnetosheath and closer to the nose of the magnetopause, while Cluster 2 (CL2) and CL4 spacecraft are in the magnetosheath with CL2 positioned above the ecliptic plane and separated from CL4 by 4.3 R\textsubscript{E}.

The magnetic field measurements by each spacecraft are shown on the right panels in Figure \ref{fig:fig2_Boview}. The Earth's bow shock recedes inward upon encountering the high dynamic pressure structure in the solar wind placing different spacecraft in the solar wind until it bounces back out. The durations of solar wind segments (gray shaded areas) are consistent with the spacecraft distance to the bow shock and the bow shock breathing motion. In the solar wind, a rotation in the magnetic field is evident in MMS and CL2 data where the $B_z$ component approaches zero. This magnetic perturbation also coincides with the second density peak of the double-peak density structure. The inward motion of the bow shock stops below the TH-A spacecraft as TH-E never crosses the bow shock or sees the solar wind. Before the end of the period, TH-E magnetic field data in Figure \ref{fig:fig2_Boview}g shows a magnetopause crossing at $\sim$ 04:03:00 UT where $B_z$ changes sign from negative in the magnetosheath to positive in the magnetosphere. The magnetic field strength however is roughly similar or slightly decreases from the magnetosheath to the magnetosphere. With the exception of  beginning and ending intervals when TH-E is in the magnetosphere, the spacecraft is in the magnetosheath region. Other spacecraft (MMS, CL2, CL4, TH-D, and TH-A) also observe the magnetosheath plasma except when excursions into the solar wind occur. 

We determine the normal vector direction during contraction and expansion motions (each side of the gray shaded areas in Figure \ref{fig:fig2_Boview}) using the minimum variance analysis \citep{sonnerup_magnetopause_1967}. The solid vectors at each spacecraft in Figure \ref{fig:fig2_Boview}a show the normal vectors during the inward motion while the dashed vectors are determined for the expanding bow shock. Based on the orientation of the normal vectors it appears that the bow shock is more curved during the expansion motion than the receding motion. The sequence of bow shock crossings also indicates that the bow shock contraction occurs faster than its expansion at Cluster and MMS orbits, while it expands faster near the nose region where THEMIS spacecraft are positioned. We compared MVA estimates of the normal vector with estimates from the mixed coplanarity method. For several crossings the two estimates are consistent and within a few degrees ($<10^{\circ}$,) while a few other estimates, particularly during the contraction phase, show larger discrepancies. Details of the MVA and mixed coplanarity analyses and spacecraft positions are listed in Table S1 in the Supplementary Material section. Normal vectors in Figure \ref{fig:fig2_Boview} are from the MVA analysis.

\subsection{Magnetic Enhancement in the Magnetosheath} \label{sec:MagEnh}
An interesting feature observed inside the magnetosheath and ahead of the receding bow shock is the magnetic enhancement observed by all THEMIS and Cluster spacecraft in Figure \ref{fig:fig2_Boview}. This structure is marked with purple boxes in magnetic field time series data. The sequence of observations in time series data begins with TH-D at 03:51:26 UT. The CL2 spacecraft observes the enhancement 4 s after TH-D, and TH-A and TH-E observe the onset of the enhancement within 9 and 12 s, respectively, after TH-D. The initial magnetic jump ratio (enhancement from the ambient magnetosheath field) is $\sim 1.3$ in all spacecraft that observe the enhancement. The magnetic field continues to grow and ratio increases to 1.4 and 1.5 at TH-E and TH-A and to $\sim 1.6$ at CL4. The MMS spacecraft does not show any increase in $|\textbf{B}|$ other than the jump associated with the bow shock crossing.  

\begin{figure}[ht]
\centering
\includegraphics[width=0.6\textwidth]{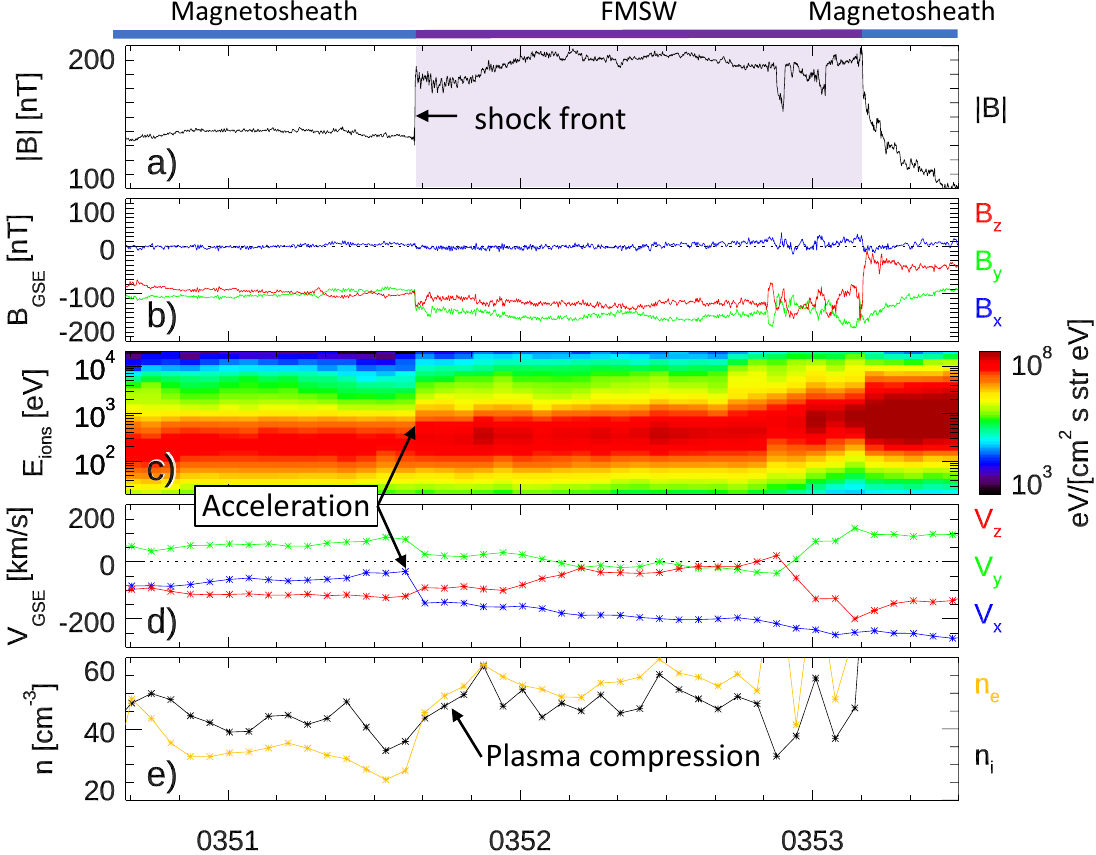}
\caption{Magnetic field and plasma measurements around the magnetic enhancement structure measured by the TH-E spacecraft. Panels (a) and (b) show the magnetic field magnitude and components. Panel (c) shows the ion energy flux spectrogram, and panels (d) and (e) show the plasma velocity and density, respectively. The fast magnetosonic shock wave (FMSW) and signatures of acceleration and compression by the shock are annotated on the figure.}
\label{fig:fig_Bcause}
\end{figure}

In Figure \ref{fig:fig_Bcause} we show the plasma and field measurements by TH-E around the magnetic peak identified by vertical dashed lines. The magnetic field components in panel (b) indicate that the field increases along the background magnetic field while the  normal vector to the shock front is mostly along the Sun-Earth line. The ion energy flux spectrogram within the magnetic enhancement in panel (c) shows higher ion energy fluxes across all energies. The flow velocity downstream of the shock is also more anti-sunward. Plasma densities in panel (e) during the magnetic enhancement also increase (both electrons and ions) by roughly similar ratios as those observed for the magnetic field enhancements. Such a correlation in density and magnetic field variations is indicative of the compressional nature of the structure. Using the mixed coplanarity method, we obtain the shock normal vector $n=[-0.90, 0.09,  0.43]$ at TH-E and $n=[-0.86,  -0.27,  0.43]$ at TH-D. The shock normal is almost exactly perpendicular to the upstream magnetic field at TH-D, while it becomes less oblique at TH-E. The propagation direction, flow velocity pattern, and compressional feature indicate that the magnetic enhancement is consistent with a fast magnetosonic shock wave (FMSW) propagating in the magnetosheath.

The magnetic enhancement ends with a sudden decrease in the absolute value of the $|B_z|$ component of the magnetic field. High ion flux intensities at around 800 eV in panel (c) after the second vertical dashed line are reminiscent of the density peak in the solar wind downstream of the bow shock. A significant plasma pileup in the magnetosheath between the magnetic structure and the receding bow shock is evident in plasma densities in the last panel of Figure \ref{fig:fig_Bcause}. The density pileup is associated with the compressed solar wind during the density peak and likely includes protons, alpha particles and He\textsuperscript{+} ions. The ion energy spectra upstream of the receding bow shock in the solar wind (around 03:54:30 UT) show the cold proton beam and additional populations of alpha particles and He\textsuperscript{+} at higher energies within the density structure. It is also worth noting that prior to the arrival of the fast magnetosonic shock wave, the magnetic field strength in the magnetosheath exhibits a gradual increase most noticeably visible in TH-E data in Figure \ref{fig:fig2_Boview}g. This magnetic field enhancement could be associated with the precursor particles from the upstream density structure entering the magnetosheath before the main density enhancement arrives. However, the sharp jump in the magnetic field corresponding to the fast shock is clearly noticeable in all THEMIS spacecraft. Such a structure not only pushes the magnetopause to further move inward, it also increases and adds to the magnetic energy density of the magnetosheath plasma near the magnetopause.

\subsection{Sunward Flows} \label{sec:sunflow}

\begin{figure}[ht]
\centering
\includegraphics[width=0.95\textwidth]{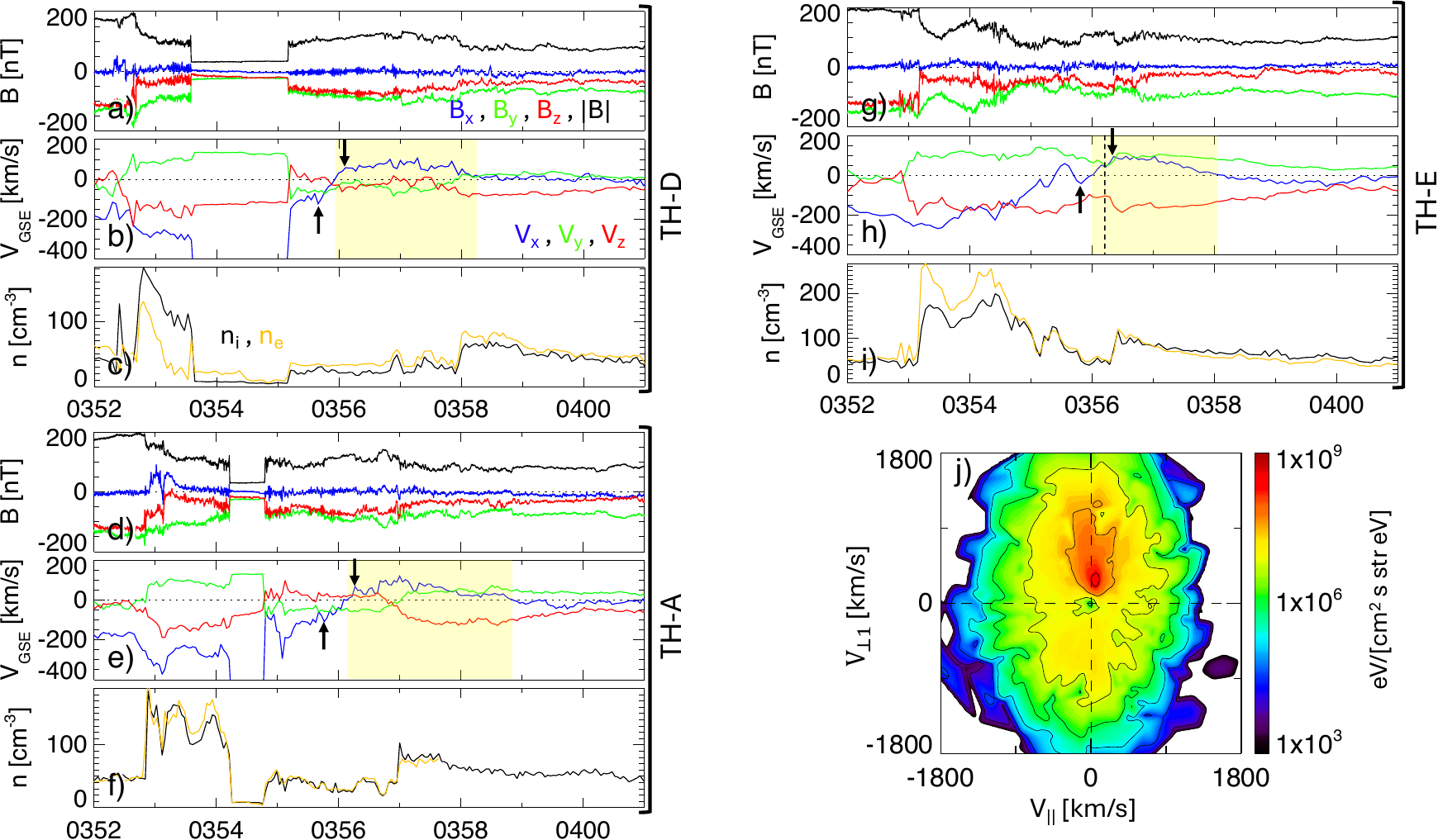}
\caption{Sunward flows in the magnetosheath plasma. Magnetic field and the plasma velocity and ion (black) and electron (gold) densities are shown from TH-D in panels (a - c), from TH-A in panels (d - f), and from TH-E in panels (g - i). Panel j shows a cut through ion distributions in the $V_{||}-V_{\perp1}$ plane at the time of the vertical dashed line in TH-E data. $V_{||}$ is parallel to the magnetic field and $V_{\perp1}$ is perpendicular to the field and along the flow velocity. The yellow shaded areas highlight the periods of sunward flows. The black arrows on velocity panels indicate data points used in linear fits to estimate the expansion rates.}
\label{fig:fig_sunwardsTHDAE}
\end{figure}

The passage of the density structure through the magnetosheath and breathing motion of the bow shock are followed by sunward plasma flows in the magnetosheath. These flows are observed by TH-D, TH-A, and TH-E spacecraft and are shown in Figure \ref{fig:fig_sunwardsTHDAE}. Measurements of the proton plasma from CL4 (not shown) do not indicate any signs of sunward flows while the MMS spacecraft near the bow shock measures anti-sunward flows \citep{madanian_interaction_2024}. Therefore, the extent of sunward flows is limited to the inner magnetosheath. The periods of sunward flows ($+V_x$) are highlighted with yellow in the velocity panels (b), (e), and (h) in Figure \ref{fig:fig_sunwardsTHDAE}. TH-D and TH-A measurements are also interrupted by an excursion into the solar wind.

The flow reversal from anti-sunward to sunward directions begins with flow becoming less anti-sunward, and the initial part of the flow reversal can be due to reduction in flux or slow down of the anti-sunward flow (i.e., transition to the fully heated solar wind plasma). As the magnetosheath plasma flow becomes sunward, the plasma densities decrease to values comparable to those observed during the fast magnetosonic shock structure (i.e., at the begining of each interval in panels (c), (f), and (i)). The maximum sunward plasma speed ($V_x$) at TH-E, TH-A, and TH-D reach as high as 94, 113, and 107 km/s, respectively. We estimate the plasma expansion acceleration rate along the Sun-Earth line in time series data using the $V_x$ component changes sign, between black arrows, and obtain sunward plasma accelerations of $dv_x/dt = 6.7$, 5.8, and 4.2 km/s\textsuperscript{2} for TH-D, TH-A and TH-E, respectively.

\begin{figure}[h]
\centering
\includegraphics[width=0.65\textwidth]{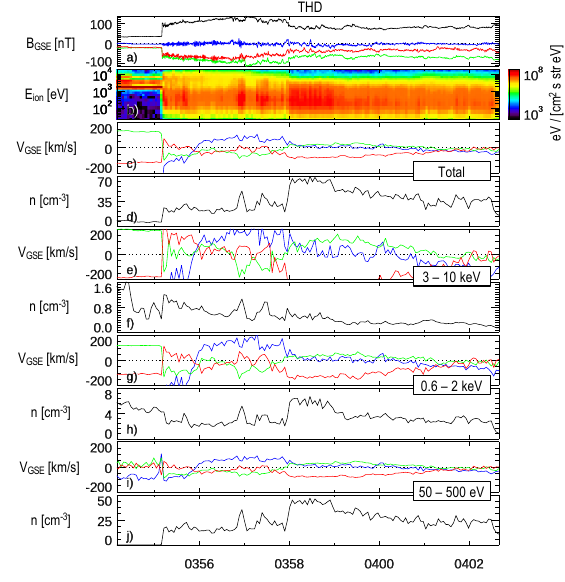}
\caption{Partial moment analysis of sunward flows observed by the THD spacecraft. Panels show: (a) magnetic field data, (b) ion energy flux spectrogram, (c-d) ion velocity and density, respectively, for all ions (of all measured energies), (e - f) 3-10 keV ions, (g – h) 0.6 – 2 keV ions, (i – j) 50 eV – 500 eV ions. The flow velocity and magnetic field components along $x$, $y$, and $z$ are shown with blue, green and red colors.}
\label{fig:fig_partial_moments}
\end{figure}

The distribution cut in Figure \ref{fig:fig_sunwardsTHDAE}j in the $BV$ plane is produced from TH-E ion distribution data (at the vertical dashed line in panel h) and shows ions at different energies in both sunward and anti-sunward directions, and parallel and anti-parallel to the magnetic field. Higher energy ions are more abundant in the perpendicular direction to the magnetic field. Determining which segment of the ion distribution constitutes the sunward flow helps in identifying the source and the underlying plasma mechanism(s). In Figure \ref{fig:fig_partial_moments} we calculate partial moments of ions at different energy ranges from TH-D ion distributions. Panels (a) and (b) show the magnetic field and energy spectra downstream of the bow shock and when sunward flows are observed. The bow shock is quasi-perpendicular. In the next two panels, the plasma velocity and density of all ions are shown (similar to Figures \ref{fig:fig_sunwardsTHDAE}b and c). The flow reversal period is evident at around 03:55:50 UT. In panels (e) and (f), we show the velocity vectors and densities of 3 keV - 10 keV ions. These ions include mostly alpha particles and singly charged helium ions that are present during this event, in addition to the high energy tail of the proton distributions. Abundances of heavy, helium group ions and the protons in the solar wind decrease near the end of the density structure (see panels f and h). Typically, in the downstream region of a quasi-perpendicular shock, ions begin to gyrate and have a velocity component perpendicular to the background magnetic field, and their guiding center pointed towards downstream. However, after the initial Earth-ward motion at the beginning of the interval, high energy ions begin to propagate sunward. Similar behavior is observed at lower energy ranges. Thus, the plasma as a whole is moving sunward while ions gyrate around the background magnetic field.

\section{Discussion} \label{sec:disc}
\subsection{Solitary Magnetic Structure in the Magnetosheath}

Unlike the observation sequence of the receding bow shock that follows the spacecraft distances from the bow shock, the first observation of the magnetosonic shock wave in the magnetosheath is by the TH-D probe which is downstream of the Cluster spacecraft. The MMS spacecraft positioned immediately downstream of the bow shock does not show such a magnetic enhancement. Therefore, it appears that certain conditions in the magnetosheath must be present for the fast magnetosonic shock to form, rather than it being launched immediately at the bow shock. The high abundances of alpha particles and singly charged helium ions during this event results in these ion populations (and also protons but to a lesser degree) to travel at super Alfvénic speeds in the magnetosheath upon crossing the bow shock \citep{madanian_interaction_2024}. If a fraction of ions crossing the bow shock remain super Alfvénic, they can cause perturbations in the magnetosheath to generate additional heating. Alpha particles and He\textsuperscript{+} ions can change the accuracy of reported plasma moments. These ions can also modify the underlying assumptions used for the fluid approximation of a planar shock \citep{lin_new_2006}.

While the magnetic field enhancement across the fast shock wave is along the background magnetic field, the wave front propagates perpendicular or at highly oblique angles to the background magnetic field. The timing analysis of the shock front observations indicates that the shock wave traveled the 0.7 R\textsubscript{E} distance along the Sun-Earth line between TH-D and TH-A spacecraft at a speed of $\sim 440$ km/s which is comparable to the fast mode wave speed in the magnetosheath ($V_f = 453$ km/s). $V_f$ increases to $\sim 550$ km/s downstream of the shock. The conservation of mass flux ($[\rho v_n]=0$) across the shock at TH-D results in a shock speed of $v_n = 241$ km/s. If we assume the difference in density between upstream and downstream of the sock is due to helium group ions, $v_n$ increases to $\sim 413$ km/s. However, doing so increases the divergence in the tangent component of the electric field across the shock. It is also worth noting that the TH-D spacecraft observes the magnetosheath shock before spacecraft upstream and downstream of it (e.g., TH-A and CL2,) indicating that the earthward propagating shock structure does not necessarily originate at the bow shock. Instead, it develops inside the magnetosheath when certain conditions are met. The shock front remains fairly unperturbed as it propagates through the inner magnetosheath, which is rare for any plasma structure in the typically turbulent magnetosheath plasma. The unperturbed nonlinear propagation pattern is suggestive of a soliton-like shock formation process. Regardless of the formation mechanism, such a magnetic enhancement increases the magnetic energy density near the magnetopause.

\subsection{Cause of Sunward Flows}
Observations of sunward flows thus far have been associated with the sunward motion of the magnetopause boundary and pressure gradient forces driving the plasma \citep{archer_global_2015, shue_anomalous_2009}. During the event discussed in this study, due to strong magnetic fields within the magnetic cloud flux rope, the ion $\beta$ in the solar wind is extremely low ($\beta <<1$), and it remains low even inside the magnetosheath (e.g., $\beta$ is $\sim 0.2$ at TH-D during the sunward flows). Thus, the plasma dynamics in the magnetosheath is dominated by the magnetic field and a thermal plasma pressure gradient cannot drive the sunward flows in the magnetosheath. Instead, the enhanced magnetic field pressure in the inner magnetosheath responds to the upstream pressure changes by creating a magnetic pressure gradient force perpendicular to the background magnetic field. Given the magnetic fields in the inner magnetosheath are dominated by $B_y$ and $B_z$ components, such a gradient force would drive the magnetosheath magnetic field lines along $x$ to expand towards rarefied upstream regions. This picture is consistent with the observations in Figures \ref{fig:fig_sunwardsTHDAE}a,d,g where higher magnetic field strengths emerge with sunward flows. In addition, the higher expansion rate observed at TH-D than at TH-A and TH-E is in agreement with a magnetosheath expansion in response to an upstream pressure decrease. During the magnetosheath sunward flows shown in Figure \ref{fig:fig_sunwardsTHDAE}, the magnetic field lines in the magnetosheath are almost entirely perpendicular to the Sun-Earth line ($B_x \sim 0$). As such, ions cannot stream along the magnetic field line to travel sunward. In addition, Figure \ref{fig:fig_partial_moments} shows that ions in all energies have a net sunward flow, including high energy ions in the 3 - 10 keV energy range. In strong magnetic fields of $\sim$ 120 nT, these ions are bound to the field lines with rather small gyroradii of $\sim 65-120$ km, and their dynamics (e.g., through reflection) cannot explain the observed sunward plasma flows in the magnetosheath \citep{fuselier_ion_1991, farrugia_effects_2018}. It seems however, the sunward plasma flows are the result of the sunward motion of the magnetic flux tubes and field lines.

\begin{figure}[h]
\centering
\includegraphics[width=0.65\textwidth]{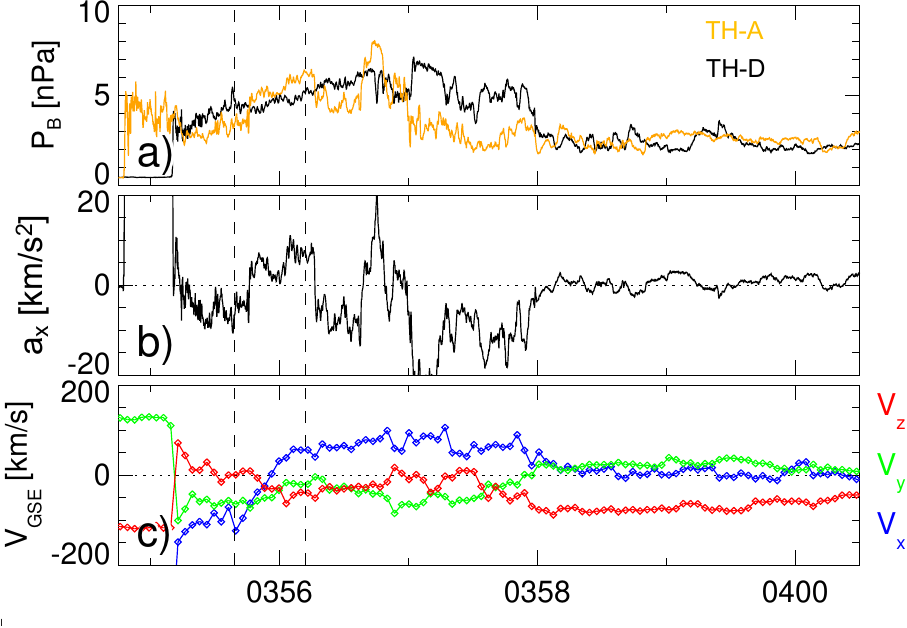}
\caption{Magnetic pressure gradient observed by THEMIS spacecraft. Panels show: (a) magnetic pressures from TH-D and TH-A, (b) plasma acceleration along the Sun-Earth line, (c) ion velocities at TH-D.}
\label{fig:fig6_pressgrad}
\end{figure}

As we discussed in Section \ref{sec:MagEnh}, the earthward propagating shock causes enhancement of the magnetic pressure in the magnetosheath. Figure \ref{fig:fig6_pressgrad}a shows magnetic pressure terms from TH-A and TH-D spacecraft. In panel (b), we estimate instantaneous acceleration rates projected along the Sun-Earth line using $a_x = \frac{1}{\rho}\nabla_x P_B$, where $\rho$ is the proton mass density at TH-D. The average of instantaneous acceleration rates between vertical dashed lines give an average acceleration rate of 2.9 km/s\textsuperscript{2}, which is comparable to the acceleration rate of 6.7 km/s\textsuperscript{2} we obtained from plasma velocities in panel (c) in the same period. This agreement between plasma acceleration and pressure gradient forcing indicates that magnetosheath sunward flows are caused by the magnetosheath expansion. Data in Figure \ref{fig:fig6_pressgrad} also indicate that after 03:58:00 UT, magnetic pressure variations in both spacecraft are decreased and the estimated acceleration rates reduce to very small values. At this time, MMS is still in the solar wind (Figure \ref{fig:fig2_Boview}b) and the magnetic pressure gradient effects are already subsided due to a short temporal scale before MMS returns back to the magnetosheath which could explain why it did not measure sunward plasma flows.

Our observations further indicate that the sunward flows are unrelated to the magnetopause boundary motion. Magnetic field measurements during the magnetopause boundary crossing at 04:03:00 UT by the TH-E spacecraft in Figure \ref{fig:fig2_Boview}g indicate that the magnetic pressure in the magnetosheath is slightly higher than the magnetospheric magnetic pressure. Thus, the magnetopause motion is limited by the high magnetic pressure in the magnetosheath and the magnetopause is controlled by the strong magnetic fields within the magnetosheath. In addition, a sunward magnetopause boundary motion leads to a more consistent flow pattern across different THEMIS spacecraft. In Figure \ref{fig:fig_sunwardsTHDAE}, when the sunward flows at TH-E begin to slow down at $\sim$ 03:57:00 UT, sunward flows at TH-D continue at the same rate and even increase at times. The observed sunward plasma flows are rather smooth and even plateau at certain velocities and are unlikely to originate at a distant reconnection zone, as we see no signs of the boundary layer plasma.

\section{Conclusions} \label{sec:conc}

In this study, we characterize a series of events in the magnetosheath caused by the interaction of an upstream density structure embedded within a strong flux rope of an ICME. Strong magnetic fields within the magnetic cloud of the ICME dominate the plasma interactions within the magnetosheath and magnetopause. We show evidence for the formation of a magnetic enhancement in the inner magnetosheath associated with a fast magnetosonic shock wave caused by the sudden surge of upstream charged particles and the associated dynamic pressure pulse. We find that the sunward flows are formed due to magnetosheath expansion and the sunward motion of the flux tubes driven by the magnetic pressure gradient force in the inner magnetosheath with sunward expansion rates as high as 6.7 km/s\textsuperscript{2}. The rarefaction effects following the density structure cause sunward flows in the magnetosheath which responds to the upstream dynamic pressure decrease by expanding sunward. These events cause significant geomagnetic activity and have significant the space weather.

\section*{Conflict of Interest Statement}
The authors declare that the research was conducted in the absence of any commercial or financial relationships that could be construed as a potential conflict of interest.

\section*{Author Contributions}
HM conceptualized the idea, performed the analysis, and prepared the initial draft of the manuscript. YP and DT contributed to interpretation of results. RR, TL, SR, TK, and JB contributed to interpretation of data. All authors contributed to preparation of the final draft of the manuscript.

\section*{Funding}
This work was supported by the NASA MMS project through the Partnership for Heliophysics and Space Environment Research (PHaSER) cooperative agreement. YP acknowledges Research Council of Finland grant number 339756. SR acknowledges funding from the MMS Early Career Award 80NSSC25K7353. RR was supported by the MMS Early Career Grant 80NSSC23K1601.

\section*{Acknowledgments}
We thank THEMIS, MMS, and Cluster operation teams for making the data available. We acknowledge the THEMIS magnetometer team at Technical University of Braunschweig for the use of FGM data, provided with the financial support by the German Federal Ministry for Economic Affairs and Climate Action (BMWK) and the German Aerospace Center (DLR) under contract 50 OC 2201. 


\section*{Data Availability Statement}
\label{sec:opres}
All data products used in this study are available through public archives at \url{https://cdaweb.gsfc.nasa.gov/}, and selecting respective missions. 

\bibliography{references}

\begin{thebibliography}{36}
\providecommand{\natexlab}[1]{#1}
\expandafter\ifx\csname urlstyle\endcsname\relax
  \providecommand{\doi}[1]{doi:\discretionary{}{}{}#1}\else
  \providecommand{\doi}{doi:\discretionary{}{}{}\begingroup \urlstyle{rm}\Url}\fi
\providecommand{\selectlanguage}[1]{\relax}
\providecommand{\bibAnnoteFile}[1]{%
  \IfFileExists{#1}{\begin{quotation}\noindent\textsc{Key:} #1\\
  \textsc{Annotation:}\ \input{#1}\end{quotation}}{}}
\providecommand{\bibAnnote}[2]{%
  \begin{quotation}\noindent\textsc{Key:} #1\\
  \textsc{Annotation:}\ #2\end{quotation}}

\bibitem[{Angelopoulos(2008)}]{angelopoulos_themis_2008}
Angelopoulos, V. (2008).
\newblock The {THEMIS} {Mission}.
\newblock \emph{Space Science Reviews} 141, 5--34.
\newblock \doi{10.1007/s11214-008-9336-1}
\bibAnnoteFile{angelopoulos_themis_2008}

\bibitem[{Archer et~al.(2015)Archer, Turner, Eastwood, Schwartz, and Horbury}]{archer_global_2015}
Archer, M., Turner, D., Eastwood, J., Schwartz, S., and Horbury, T. (2015).
\newblock Global impacts of a {Foreshock} {Bubble}: {Magnetosheath}, magnetopause and ground-based observations.
\newblock \emph{Planetary and Space Science} 106, 56--66.
\newblock \doi{10.1016/j.pss.2014.11.026}
\bibAnnoteFile{archer_global_2015}

\bibitem[{Archer et~al.(2014)Archer, Turner, Eastwood, Horbury, and Schwartz}]{archer_role_2014}
Archer, M.~O., Turner, D.~L., Eastwood, J.~P., Horbury, T.~S., and Schwartz, S.~J. (2014).
\newblock The role of pressure gradients in driving sunward magnetosheath flows and magnetopause motion.
\newblock \emph{Journal of Geophysical Research: Space Physics} 119, 8117--8125.
\newblock \doi{10.1002/2014JA020342}
\bibAnnoteFile{archer_role_2014}

\bibitem[{Broll et~al.(2018)Broll, Fuselier, Trattner, Schwartz, Burch, Giles et~al.}]{broll_mms_2018}
Broll, J.~M., Fuselier, S.~A., Trattner, K.~J., Schwartz, S.~J., Burch, J.~L., Giles, B.~L., et~al. (2018).
\newblock {MMS} {Observation} of {Shock}-{Reflected} {He} $^{\textrm{++}}$ at {Earth}'s {Quasi}-{Perpendicular} {Bow} {Shock}.
\newblock \emph{Geophysical Research Letters} 45, 49--55.
\newblock \doi{10.1002/2017GL075411}
\bibAnnoteFile{broll_mms_2018}

\bibitem[{Burch et~al.(2016)Burch, Moore, Torbert, and Giles}]{burch_magnetospheric_2016}
Burch, J.~L., Moore, T.~E., Torbert, R.~B., and Giles, B.~L. (2016).
\newblock Magnetospheric {Multiscale} {Overview} and {Science} {Objectives}.
\newblock \emph{Space Science Reviews} 199, 5--21.
\newblock \doi{10.1007/s11214-015-0164-9}
\bibAnnoteFile{burch_magnetospheric_2016}

\bibitem[{Burgess(1989)}]{burgess_alpha_1989}
Burgess, D. (1989).
\newblock Alpha particles in field-aligned beams upstream of the bow shock: {Simulations}.
\newblock \emph{Geophysical Research Letters} 16.
\newblock \doi{10.1029/GL016i002p00163}
\bibAnnoteFile{burgess_alpha_1989}

\bibitem[{Burgess et~al.(2016)Burgess, Hellinger, Gingell, and Trávníček}]{burgess_microstructure_2016}
Burgess, D., Hellinger, P., Gingell, I., and Trávníček, P.~M. (2016).
\newblock Microstructure in two- and three-dimensional hybrid simulations of perpendicular collisionless shocks.
\newblock \emph{Journal of Plasma Physics} 82, 905820401.
\newblock \doi{10.1017/S0022377816000660}
\bibAnnoteFile{burgess_microstructure_2016}

\bibitem[{Burkholder et~al.(2023)Burkholder, Chen, Sorathia, Sciola, Merkin, Trattner et~al.}]{burkholder_complexity_2023}
Burkholder, B.~L., Chen, L.-J., Sorathia, K., Sciola, A., Merkin, S., Trattner, K.~J., et~al. (2023).
\newblock The complexity of the day-side {X}-line during southward interplanetary magnetic field.
\newblock \emph{Frontiers in Astronomy and Space Sciences} 10, 1175697.
\newblock \doi{10.3389/fspas.2023.1175697}
\bibAnnoteFile{burkholder_complexity_2023}

\bibitem[{Chapman and Ferraro(1931)}]{chapman_new_1931}
Chapman, S. and Ferraro, V. C.~A. (1931).
\newblock A new theory of magnetic storms.
\newblock \emph{Terrestrial Magnetism and Atmospheric Electricity} 36, 77--97.
\newblock \doi{10.1029/TE036i002p00077}
\bibAnnoteFile{chapman_new_1931}

\bibitem[{Escoubet et~al.(2001)Escoubet, Fehringer, and Goldstein}]{escoubet_cluster_2001}
Escoubet, C.~P., Fehringer, M., and Goldstein, M. (2001).
\newblock The {Cluster} mission.
\newblock \emph{Annales Geophysicae} 19, 1197--1200.
\newblock \doi{10.5194/angeo-19-1197-2001}
\bibAnnoteFile{escoubet_cluster_2001}

\bibitem[{Farris and Russell(1994)}]{farris_determining_1994}
Farris, M.~H. and Russell, C.~T. (1994).
\newblock Determining the standoff distance of the bow shock: {Mach} number dependence and use of models.
\newblock \emph{Journal of Geophysical Research} 99, 17681.
\newblock \doi{10.1029/94JA01020}
\bibAnnoteFile{farris_determining_1994}

\bibitem[{Farrugia et~al.(2018)Farrugia, Cohen, Vasquez, Lugaz, Alm, Torbert et~al.}]{farrugia_effects_2018}
Farrugia, C.~J., Cohen, I.~J., Vasquez, B.~J., Lugaz, N., Alm, L., Torbert, R.~B., et~al. (2018).
\newblock Effects in the {Near}‐{Magnetopause} {Magnetosheath} {Elicited} by {Large}‐{Amplitude} {Alfvénic} {Fluctuations} {Terminating} in a {Field} and {Flow} {Discontinuity}.
\newblock \emph{Journal of Geophysical Research: Space Physics} 123, 8983--9004.
\newblock \doi{10.1029/2018JA025724}
\bibAnnoteFile{farrugia_effects_2018}

\bibitem[{Fatemi et~al.(2024)Fatemi, Hamrin, Krämer, Gunell, Nordin, Karlsson et~al.}]{fatemi_unveiling_2024}
Fatemi, S., Hamrin, M., Krämer, E., Gunell, H., Nordin, G., Karlsson, T., et~al. (2024).
\newblock Unveiling the {3D} structure of magnetosheath jets.
\newblock \emph{Monthly Notices of the Royal Astronomical Society} 531, 4692--4713.
\newblock \doi{10.1093/mnras/stae1456}
\bibAnnoteFile{fatemi_unveiling_2024}

\bibitem[{Fuselier et~al.(1991)Fuselier, Klumpar, and Shelley}]{fuselier_ion_1991}
Fuselier, S.~A., Klumpar, D.~M., and Shelley, E.~G. (1991).
\newblock Ion {Reflection} and transmission during reconnection at the {Earth}'s subsolar magnetopause.
\newblock \emph{Geophysical Research Letters} 18, 139--142.
\newblock \doi{10.1029/90GL02676}
\bibAnnoteFile{fuselier_ion_1991}

\bibitem[{Gurchumelia et~al.(2022)Gurchumelia, Sorriso-Valvo, Burgess, Yordanova, Elbakidze, Kharshiladze et~al.}]{gurchumelia_comparing_2022}
Gurchumelia, A., Sorriso-Valvo, L., Burgess, D., Yordanova, E., Elbakidze, K., Kharshiladze, O., et~al. (2022).
\newblock Comparing {Quasi}-{Parallel} and {Quasi}-{Perpendicular} {Configuration} in the {Terrestrial} {Magnetosheath}: {Multifractal} {Analysis}.
\newblock \emph{Frontiers in Physics} 10, 903632.
\newblock \doi{10.3389/fphy.2022.903632}
\bibAnnoteFile{gurchumelia_comparing_2022}

\bibitem[{Harten and Clark(1995)}]{harten_design_1995}
Harten, R. and Clark, K. (1995).
\newblock The design features of the {GGS} wind and polar spacecraft.
\newblock \emph{Space Science Reviews} 71, 23--40.
\newblock \doi{10.1007/BF00751324}
\bibAnnoteFile{harten_design_1995}

\bibitem[{Karlsson et~al.(2015)Karlsson, Kullen, Liljeblad, Brenning, Nilsson, Gunell et~al.}]{karlsson_origin_2015}
Karlsson, T., Kullen, A., Liljeblad, E., Brenning, N., Nilsson, H., Gunell, H., et~al. (2015).
\newblock On the origin of magnetosheath plasmoids and their relation to magnetosheath jets: {ON} {THE} {ORIGIN} {OF} {MAGNETOSHEATH} {PLASMOIDS}.
\newblock \emph{Journal of Geophysical Research: Space Physics} 120, 7390--7403.
\newblock \doi{10.1002/2015JA021487}
\bibAnnoteFile{karlsson_origin_2015}

\bibitem[{Krasnoselskikh et~al.(2013)Krasnoselskikh, Balikhin, Walker, Schwartz, Sundkvist, Lobzin et~al.}]{krasnoselskikh_dynamic_2013}
Krasnoselskikh, V., Balikhin, M., Walker, S.~N., Schwartz, S., Sundkvist, D., Lobzin, V., et~al. (2013).
\newblock The dynamic quasiperpendicular shock: {Cluster} discoveries.
\newblock \emph{Space Science Reviews} 178, 535--598.
\newblock \doi{10.1007/s11214-013-9972-y}.
\newblock ArXiv: 1303.0190
\bibAnnoteFile{krasnoselskikh_dynamic_2013}

\bibitem[{Krämer et~al.(2025)Krämer, Koller, Suni, LaMoury, Pöppelwerth, Glebe et~al.}]{kramer_jets_2025}
Krämer, E., Koller, F., Suni, J., LaMoury, A.~T., Pöppelwerth, A., Glebe, G., et~al. (2025).
\newblock Jets {Downstream} of {Collisionless} {Shocks}: {Recent} {Discoveries} and {Challenges}.
\newblock \emph{Space Science Reviews} 221, 4.
\newblock \doi{10.1007/s11214-024-01129-3}
\bibAnnoteFile{kramer_jets_2025}

\bibitem[{Lin et~al.(2006)Lin, Chao, Lee, Lyu, and Wu}]{lin_new_2006}
Lin, C.~C., Chao, J.~K., Lee, L.~C., Lyu, L.~H., and Wu, D.~J. (2006).
\newblock A new shock fitting procedure for the {MHD} {Rankine}‐{Hugoniot} relations for the case of small {He}$^{\textrm{2+}}$ slippage.
\newblock \emph{Journal of Geophysical Research: Space Physics} 111, 2005JA011449.
\newblock \doi{10.1029/2005JA011449}
\bibAnnoteFile{lin_new_2006}

\bibitem[{Lin et~al.(1995)Lin, Anderson, Ashford, Carlson, Curtis, Ergun et~al.}]{lin_three-dimensional_1995}
Lin, R.~P., Anderson, K.~A., Ashford, S., Carlson, C., Curtis, D., Ergun, R., et~al. (1995).
\newblock A three-dimensional plasma and energetic particle investigation for the wind spacecraft.
\newblock \emph{Space Science Reviews} 71, 125--153.
\newblock \doi{10.1007/BF00751328}
\bibAnnoteFile{lin_three-dimensional_1995}

\bibitem[{Liu et~al.(2024)Liu, Shi, Hartinger, Angelopoulos, Rodger, Viljanen et~al.}]{liu_global_2024}
Liu, T.~Z., Shi, X., Hartinger, M.~D., Angelopoulos, V., Rodger, C.~J., Viljanen, A., et~al. (2024).
\newblock Global {Observations} of {Geomagnetically} {Induced} {Currents} {Caused} by an {Extremely} {Intense} {Density} {Pulse} {During} a {Coronal} {Mass} {Ejection}.
\newblock \emph{Space Weather} 22, e2024SW003993.
\newblock \doi{10.1029/2024SW003993}
\bibAnnoteFile{liu_global_2024}

\bibitem[{Madanian et~al.(2024{\natexlab{a}})Madanian, Chen, Ng, Starkey, Fuselier, Bessho et~al.}]{madanian_interaction_2024}
Madanian, H., Chen, L.-J., Ng, J., Starkey, M.~J., Fuselier, S.~A., Bessho, N., et~al. (2024{\natexlab{a}}).
\newblock Interaction of the {Prominence} {Plasma} within the {Magnetic} {Cloud} of an {Interplanetary} {Coronal} {Mass} {Ejection} with the {Earth}’s {Bow} {Shock}.
\newblock \emph{The Astrophysical Journal} 976, 219.
\newblock \doi{10.3847/1538-4357/ad8579}
\bibAnnoteFile{madanian_interaction_2024}

\bibitem[{Madanian et~al.(2024{\natexlab{b}})Madanian, Gingell, Chen, and Monyek}]{madanian_drivers_2024}
Madanian, H., Gingell, I., Chen, L.-J., and Monyek, E. (2024{\natexlab{b}}).
\newblock Drivers of {Magnetic} {Field} {Amplification} at {Oblique} {Shocks}: {In} {Situ} {Observations}.
\newblock \emph{The Astrophysical Journal Letters} 965, L12.
\newblock \doi{10.3847/2041-8213/ad3073}
\bibAnnoteFile{madanian_drivers_2024}

\bibitem[{Madanian et~al.(2022)Madanian, Liu, Phan, Trattner, Karlsson, and Liemohn}]{madanian_asymmetric_2022}
Madanian, H., Liu, T.~Z., Phan, T.~D., Trattner, K.~J., Karlsson, T., and Liemohn, M.~W. (2022).
\newblock Asymmetric {Interaction} of a {Solar} {Wind} {Reconnecting} {Current} {Sheet} and {Its} {Magnetic} {Hole} {With} {Earth}'s {Bow} {Shock} and {Magnetopause}.
\newblock \emph{Journal of Geophysical Research: Space Physics} 127.
\newblock \doi{10.1029/2021JA030079}
\bibAnnoteFile{madanian_asymmetric_2022}

\bibitem[{Maynard et~al.(2008)Maynard, Farrugia, Ober, Burke, Dunlop, Mozer et~al.}]{maynard_cluster_2008}
Maynard, N.~C., Farrugia, C.~J., Ober, D.~M., Burke, W.~J., Dunlop, M., Mozer, F.~S., et~al. (2008).
\newblock Cluster observations of fast shocks in the magnetosheath launched as a tangential discontinuity with a pressure increase crossed the bow shock: {FAST} {SHOCKS} {IN} {THE} {MAGNETOSHEATH}.
\newblock \emph{Journal of Geophysical Research: Space Physics} 113.
\newblock \doi{10.1029/2008JA013121}
\bibAnnoteFile{maynard_cluster_2008}

\bibitem[{Phan et~al.(2010)Phan, Gosling, Paschmann, Pasma, Drake, Øieroset et~al.}]{phan_dependence_2010}
Phan, T.~D., Gosling, J.~T., Paschmann, G., Pasma, C., Drake, J.~F., Øieroset, M., et~al. (2010).
\newblock {THE} {DEPENDENCE} {OF} {MAGNETIC} {RECONNECTION} {ON} {PLASMA} {B} {AND} {MAGNETIC} {SHEAR}: {EVIDENCE} {FROM} {SOLAR} {WIND} {OBSERVATIONS}.
\newblock \emph{The Astrophysical Journal} 719, L199--L203.
\newblock \doi{10.1088/2041-8205/719/2/L199}
\bibAnnoteFile{phan_dependence_2010}

\bibitem[{Plaschke et~al.(2013)Plaschke, Angelopoulos, and Glassmeier}]{plaschke_magnetopause_2013}
Plaschke, F., Angelopoulos, V., and Glassmeier, K. (2013).
\newblock Magnetopause surface waves: {THEMIS} observations compared to {MHD} theory.
\newblock \emph{Journal of Geophysical Research: Space Physics} 118, 1483--1499.
\newblock \doi{10.1002/jgra.50147}
\bibAnnoteFile{plaschke_magnetopause_2013}

\bibitem[{Plaschke et~al.(2018)Plaschke, Hietala, Archer, Blanco-Cano, Kajdič, Karlsson et~al.}]{plaschke_jets_2018}
Plaschke, F., Hietala, H., Archer, M., Blanco-Cano, X., Kajdič, P., Karlsson, T., et~al. (2018).
\newblock Jets {Downstream} of {Collisionless} {Shocks}.
\newblock \emph{Space Science Reviews} 214.
\newblock \doi{10.1007/s11214-018-0516-3}
\bibAnnoteFile{plaschke_jets_2018}

\bibitem[{Schwartz et~al.(2022)Schwartz, Goodrich, Wilson, Turner, Trattner, Kucharek et~al.}]{schwartz_energy_2022}
Schwartz, S.~J., Goodrich, K.~A., Wilson, L.~B., Turner, D.~L., Trattner, K.~J., Kucharek, H., et~al. (2022).
\newblock Energy {Partition} at {Collisionless} {Supercritical} {Quasi}‐{Perpendicular} {Shocks}.
\newblock \emph{Journal of Geophysical Research: Space Physics} 127, e2022JA030637.
\newblock \doi{10.1029/2022JA030637}
\bibAnnoteFile{schwartz_energy_2022}

\bibitem[{Shue et~al.(2009)Shue, Chao, Song, McFadden, Suvorova, Angelopoulos et~al.}]{shue_anomalous_2009}
Shue, J., Chao, J., Song, P., McFadden, J.~P., Suvorova, A., Angelopoulos, V., et~al. (2009).
\newblock Anomalous magnetosheath flows and distorted subsolar magnetopause for radial interplanetary magnetic fields.
\newblock \emph{Geophysical Research Letters} 36, 2009GL039842.
\newblock \doi{10.1029/2009GL039842}
\bibAnnoteFile{shue_anomalous_2009}

\bibitem[{Shue et~al.(1998)Shue, Song, Russell, Steinberg, Chao, Zastenker et~al.}]{shue_magnetopause_1998}
Shue, J.-H., Song, P., Russell, C.~T., Steinberg, J.~T., Chao, J.~K., Zastenker, G., et~al. (1998).
\newblock Magnetopause location under extreme solar wind conditions.
\newblock \emph{Journal of Geophysical Research: Space Physics} 103, 17691--17700.
\newblock \doi{10.1029/98JA01103}
\bibAnnoteFile{shue_magnetopause_1998}

\bibitem[{Siscoe et~al.(1980)Siscoe, Crooker, and Belcher}]{siscoe_sunward_1980}
Siscoe, G.~L., Crooker, N.~U., and Belcher, J.~W. (1980).
\newblock Sunward flow in {Jupiter}'s magnetosheath.
\newblock \emph{Geophysical Research Letters} 7, 25--28.
\newblock \doi{10.1029/GL007i001p00025}
\bibAnnoteFile{siscoe_sunward_1980}

\bibitem[{Sonnerup and Cahill(1967)}]{sonnerup_magnetopause_1967}
Sonnerup, B.~U. and Cahill, L.~J. (1967).
\newblock Magnetopause structure and attitude from {Explorer} 12 observations.
\newblock \emph{Journal of Geophysical Research} 72, 171.
\newblock \doi{10.1029/JZ072i001p00171}
\bibAnnoteFile{sonnerup_magnetopause_1967}

\bibitem[{Wu et~al.(1993)Wu, Mandt, Lee, and Chao}]{wu_magnetospheric_1993}
Wu, B.~H., Mandt, M.~E., Lee, L.~C., and Chao, J.~K. (1993).
\newblock Magnetospheric {Response} to {Solar} {Wind} {Dynamic} {Pressure} {Variations}: {Interaction} of {Interplanetary} {Tangential} {Discontinuities} with the {Bow} {Shock}.
\newblock \emph{Journal of Geophysical Research} 98, 21297--21311.
\newblock \doi{10.1029/93JA01013}
\bibAnnoteFile{wu_magnetospheric_1993}

\bibitem[{Zhou et~al.(2024)Zhou, Raptis, Wang, Shen, Ren, and Ma}]{zhou_magnetosheath_2024}
Zhou, Y., Raptis, S., Wang, S., Shen, C., Ren, N., and Ma, L. (2024).
\newblock Magnetosheath jets at {Jupiter} and across the solar system.
\newblock \emph{Nature Communications} 15, 4.
\newblock \doi{10.1038/s41467-023-43942-4}
\bibAnnoteFile{zhou_magnetosheath_2024}

\end{thebibliography}
\bibliographystyle{Frontiers-Harvard}

\end{document}